\documentclass[a4paper,11pt]{article}

\usepackage{pos}

\title{Highlights from the IceCube Neutrino Observatory}

\ShortTitle{Highlights from the IceCube Neutrino Observatory}

\author{The IceCube Collaboration \\{\normalsize \normalfont(a complete list of authors can be found at the end of the proceedings)}\\}

\emailAdd{naoko@drexel.edu}

\abstract{

As IceCube surpasses a decade of operation in the full detector configuration, results that drive forward the fields of neutrino astronomy, cosmic ray physics, multi-messenger astronomy, particle physics, and beyond continue to emerge at an accelerated pace. IceCube data is dominated by background events, and thus teasing out the signal is the common challenge to most analyses. Statistical accumulation of data, along with better understanding of the background fluxes, the detector, and continued development of our analysis tools have produced many profound results that were presented at ICRC2023. Highlights covered here include the first neutrino observation of the Galactic Plane, the first observation of a steady emission neutrino point source NGC1068, new characterizations of the cosmic ray flux and its secondary particles, and a possible new era in measuring the energy spectrum of the diffuse astrophysical flux. IceCube is poised to make more discoveries and drive fields forward in the near future with many novel analyses coming online.

\vspace{4mm}
{\bfseries Corresponding authors:}
Naoko Kurahashi$^{1*}$\\
{$^{1}$ \itshape Department of Physics, Drexel University, Philadelphia, Pennsylvania, USA}\\[4mm]
$^*$ Presenter

\ConferenceLogo{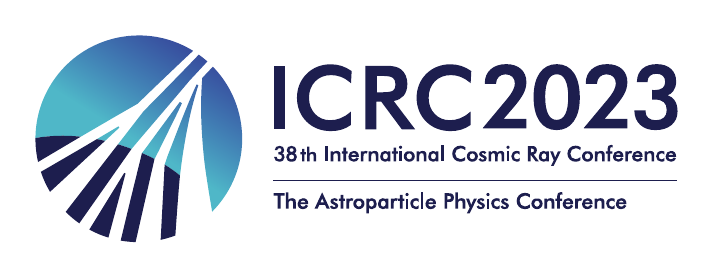}

\FullConference{The 38th International Cosmic Ray Conference (ICRC2023)\\ 26 July -- 3 August, 2023\\ Nagoya, Japan}
}

\begin{document}

\maketitle

\section{Introduction}\label{sec1}
The IceCube Neutrino Observatory is a cubic-kilometer-scale neutrino detector located at the geographic South Pole~\cite{icecube_detector}. While data taking was ongoing during construction, the full detector was completed in 2011, and it has now accumulated over a decade of full-detector data. As data accumulates and statistical errors shrink, IceCube's many analysis channels are starting to probe paradigm-shifting regions of sensitivity. Profound conclusions in different subfields are emerging from IceCube. ICRC2023 highlighted many of these results. Here we present an overview of selected results. However, details for each analysis can be found in the cited proceedings, and all IceCube contributions can be found in Ref.~\cite{Author:2023icrc}.

\section{IceCube Data}
IceCube observes neutrinos by measuring Cherenkov light in the detector emitted by charged secondary particles as a result of neutrinos interacting with the glacial ice and the antarctic bedrock. This measurement is used to reconstruct the direction and energy of the primary neutrinos. The light patterns observed in the detector can be categorized into two main topologies.  ``Track'' events are muons that traverse the detector producing a straight line of light, since high-energy muons are long-lived with relatively small energy loss. Muons are produced by  ``charged-current'' (CC) $\nu_\mu$ interactions. The other topology is  ``cascade'' events which are particle showers induced by ``neutral-current'' (NC) interactions of all flavors, as well as CC$-\nu_e$ or CC-$\nu_\tau$ interactions. Because of the very short shower length compared to the detector size, and the optical properties of the glacial ice with short scattering distances but long absorption lengths, the light emitted from these showers manifests as nearly spherically expanding.  For this reason, typical track events have superior angular resolution when compared to cascade events, but most cascade events have superior energy resolution as a larger fraction of energy loss is observed in the detector. 

For every astrophysical neutrino event observed in the detector, there are around 3 orders of magnitude more atmospheric neutrino events observed at trigger level. For every atmospheric neutrino event, there are around 6 orders of magnitude more atmospheric muon events observed. Unless one is performing an analysis that aims to understand a feature seen in all of the atmospheric muons, e.g., a study of cosmic ray anisotropy~\cite{Author:2023icrcCRAnisotropy}, the IceCube data will be background dominated. Thus in almost all IceCube analyses, regardless of the subfield area of study they belong to or the method of the analysis, combating background events is the largest challenge. Depending on the goal of the analysis, many techniques are used, including cut and characterize, cut and scramble, employ harsh cuts, or loosen the cuts and fit the backgrounds. Either way, understanding our data, the statistical properties and the systematic effects, become key to a successful IceCube analysis. 

\section{Galactic Neutrino Astronomy}\label{sec2}

\begin{figure}
    \centering
    \includegraphics[width=\textwidth]{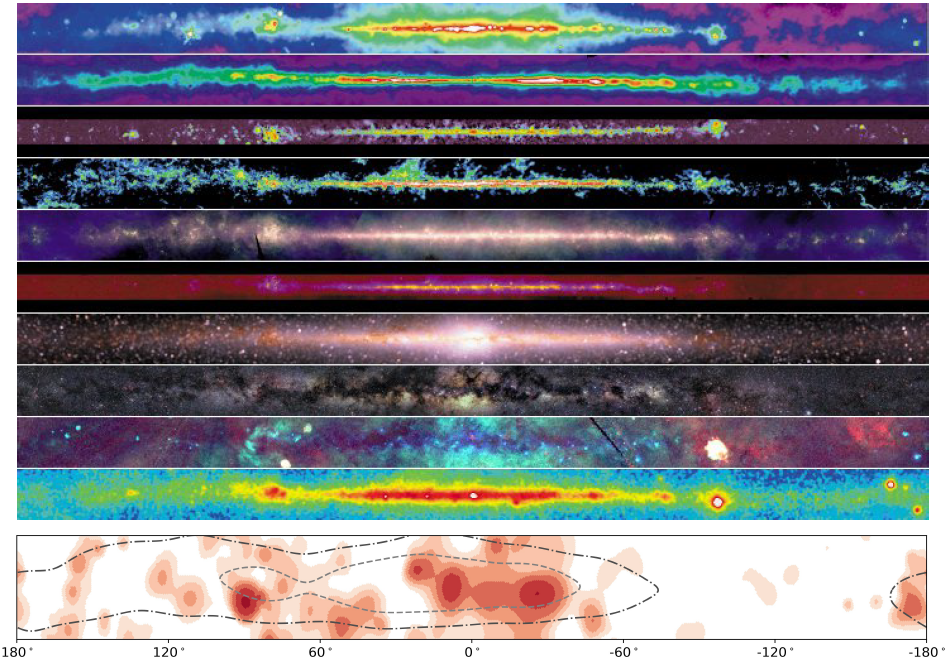}
    \caption{Iconic image of the multiwavelength view of the Milky Way. Galactic latitude of $\pm 10^{o}$ is shown from radio to gamma ray observations~\cite{multiwave_GP}. At the bottom is the "neutrino image" of the Galactic plane shown for latitudes of $\pm 20^{o}$}
    \label{fig:gp}
\end{figure}

IceCube presented the first observation of the Milky Way Galaxy in neutrinos at a statistical significance 4.5~$\sigma$ at this conference~\cite{Author:2023icrcGP}, as was previously announced in Ref.~\cite{sciencearticle}. High energy neutrinos are produced when cosmic rays interact at their acceleration sites and during propagation through the interstellar medium. The Galactic plane has therefore long been hypothesized as a neutrino source. IceCube's observation of the diffuse Galactic Plane represents the first non-electromagnetic image of our galaxy\footnote{With much respect to our solar neutrino colleagues who have been observing the sun in MeV for decades, and more recently in lower energies. Here we mean the Galaxy as a whole.}.

Because the Galactic center is located in the Southern Sky, diffuse emission is  expected to be concentrated in the Southern Sky.  IceCube is located at the South Pole, so observations in the southern celestial sky are composed of events downgoing in the detector.  Searches in this region are particularly difficult due to the large background of atmospheric muons. Through-going track-based analyses see a reduction of sensitivity due to this irreducible background~\cite{tracks10yr}.  This is especially true for galactic sources, which are assumed to follow a softer spectrum than extragalactic fluxes. The observation of the Galactic Plane was only made possible because it makes use of cascade events, which are usually not used in source searches due to the larger angular resolution. However cascade events have a much reduced background in the southern sky, and despite their angular resolution, this leads to an improvement of sensitivity.  Further, machine learning techniques applied to IceCube cascade events improve the sensitivity.

Many questions arise from the first observation of the Galactic Plane in neutrinos. One main question is the contribution of yet unknown but future-resolvable point sources to the diffuse emission. Answering this will most likely require track events~\cite{Author:2023icrcGS}. However, a simultaneous spectral fit of the diffuse Galactic and Extragalactic fluxes has so far not yielded a statistically significant presence of the Galactic flux above 3$\sigma$ in the track data set~\cite{Author:2023icrcNorthTrack}. 

With the Galactic Plane observed, emission from the Galactic Center (GC) also becomes interesting. The GC is a region that promises high activity based on the detection of a PeVatron in the GC and the presence of a supermassive black hole (SMBH) at the position of Sgr A*. SMBHs can be sources of flare-like emission of cosmic rays and their secondaries, neutrinos and gamma-rays. This motivates our search for a single flare from the GC. Through the utilization of both conventional and machine learning techniques, a new event selection is conducted on IceCube data within the GC region. Employing this new dataset, a search for time-dependent single flares in the GC region has been executed, however, no significant flares were found~\cite{Author:2023icrcGP}.

While the era of Galatic high-energy neutrino astronomy has commenced with the observation of the diffuse plane, identifying individual Galactic sources remains a challenge. Detectors ongoing construction and in future plans, such as IceCube-Gen2~\cite{Author:2023icrcGen2}, KM3NeT~\cite{Author:2023icrcKM3NeT}, P-ONE~\cite{Author:2023icrcPONE}, and others may provide a more complete answer. A summary of the contributions by the IceCube-Gen2 project can be found in Ref.~\cite{Author:2023icrcGen2Bundle}.

\section{Energy Spectra of Diffuse Neutrino Fluxes}\label{sec3}
Since the first observation of the all-sky diffuse astrophysical neutrino spectrum in 2013~\cite{sciencearticle}, a steady statistical increase in accumulated data has allowed many observation channels to measure this flux at high statistical significance. The first observation used the channel of High Energy Starting Events (HESE). The most recent iteration of the HESE analysis, which is sensitive to neutrinos above 60~TeV, measures the astrophysical flux with energy spectrum consistent with a single power law spectrum with best-fit index $2.87^{+0.20}_{-0.19}$~\cite{HESE_paper}. This is softer than other IceCube measurements of the astrophysical neutrino spectrum. Extending this technique down to 1~TeV energies is the Medium Energy Starting Events (MESE) channel. Plans to include a more accurate modeling of the detector self-veto, along with a larger sample size to investigate a possible excess at 30~TeV seen with a previous 2-year MESE data set~\cite{MESE_paper} was presented~\cite{Author:2023icrcMESE}.

Combining previous diffuse flux measurements that include both track and cascade channels, with consistent nuisance parameter treatment, is the so called "global" fit. The result~\cite{Author:2023icrcGlobalFit} sees a possible deviation from a power law at the highest energies (>few 100 TeV reconstructed energies), as well as more possible indications for structure in the astrophysical flux, most prominent in the excess around 30 TeV reconstructed energy. This may indicate a deviation from the single power law expectation, since a preference for a break in the spectrum is observed.

Furthermore, new channels that are only now becoming available with the accumulation of high statistics data are observing the astrophysical flux too. One such channel is the starting track events, which is not part of the global fit. The starting track events analysis~\cite{Author:2023icrcEstes} marks the first time IceCube has measured the astrophysical diffuse flux using a data set composed primarily of starting track events. Starting tracks combine the superior angular resolution of tracks and energy resolution of cascades. This statistically challenging data set takes advantage of the self-veto effect in the Southern Sky reducing the atmospheric neutrino background. The best fit single power law index of the astrophysical neutrino flux is $2.58^{+0.10}_{-0.09}$. The astrophysical flux $90\%$ sensitive energy range is 3 TeV to 500 TeV, extending IceCube's reach to the low energy astrophysical flux by an order of magnitude. 

\begin{figure}
    \centering
    \includegraphics[width=\textwidth]{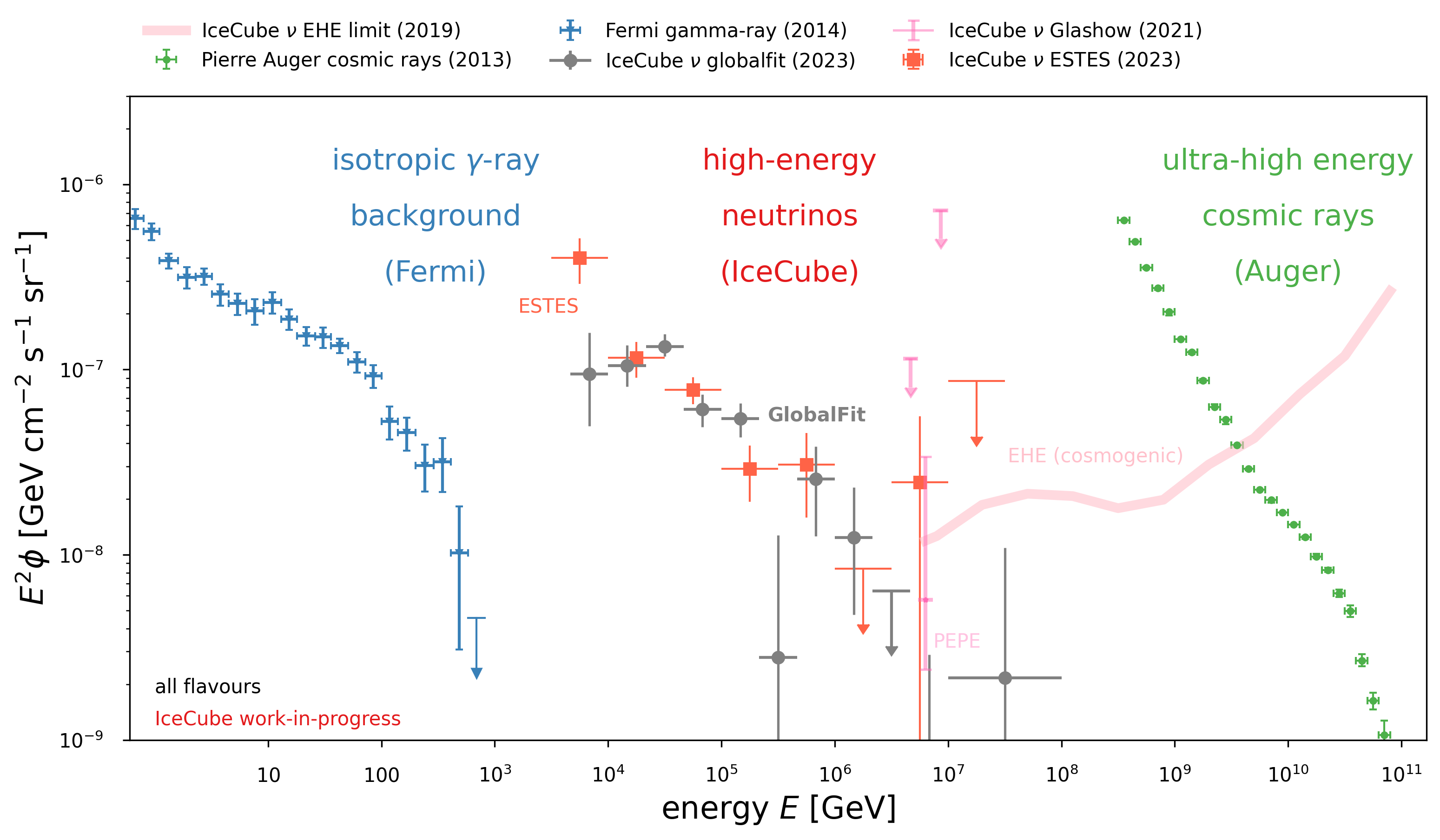}
    \caption{The measured energy spectra of diffuse astrophysical flux by the global fit analysis and the starting track events analysis (labeled "ESTES"), along with other observed diffuse fluxes.}
    \label{fig:diffuse}
\end{figure}

At the discovery of the astrophysical diffuse flux, an all-sky single component power law needs to be assumed in order to fit the entire range of the energy spectrum, due to the limited event statistics available above the background events that dominate. Now with growing data sets and growing detection channels, it is possible that we are at a point where differences in regions of the sky, energy ranges, and possibly even neutrino flavors can be taken into account. Fig.~\ref{fig:diffuse} shows the comparison between the global fit and the new starting track channel, along with other measurements. The next few years may become the turning point in which more parameters of the diffuse flux are measured with high statistical certainty. 

Tau neutrinos, which have unique signatures of double cascades where a tau neutrino interacts with the glacial ice's nucleon to produce a tau lepton that then travels some distance in ice and decays into an electron or multiple hadrons,  can shed further light on the possibly different diffuse astrophysical fluxes for different neutrino flavors. This is another example of a new channel that is only now available with the accumulation of large data. Seven candidate events were found in 10 years of data, consistent with the 1:1:1 flavor ratio of diffuse astrophysical neutrinos~\cite{Author:2023icrcTau}.

Beyond the astrophysical diffuse flux, at higher energies, lies an expected flux of cosmic neutrinos, or GZK neutrinos, produced when ultra high energy cosmic rays interact with ambient photons of the cosmic microwave background. An expanded analysis plan aimed to observe these extremely high energy neutrinos was presented~\cite{Author:2023icrcGZK}.

Finally, using the astrophysical diffuse flux, a particle physics phenomenon predicted over 60 years ago has been observed for the first time. The Glashow resonance is the resonant formation of a W$-$ boson during the interaction of a high-energy electron antineutrino with an electron, peaking at an antineutrino energy of 6.3~PeV in the rest frame of the electron. Whereas this energy scale is out of reach still at particle accelerators, the astrophysical diffuse flux provides high energy anti-neutrinos at this energy. The expected flux level of astrophysical electron antineutrinos at such high energies is extremely low, but due to this effect the cross section enhances dramatically. A detection by the IceCube neutrino observatory of a cascade event consistent with being created at the Glashow resonance has recently been published~\cite{IceCube:Glashow}.

\section{Cosmic Rays and Atmospheric Neutrinos}\label{sec4}
IceCube probes the cosmic ray flux in several ways. The IceTop detector~\cite{icetop_paper}, consisting of 81 pairs of Ice-Cherenkov tanks, placed on the surface of the antarctic ice above the in-ice part of the IceCube detector, covers an area of $1~\mathrm{km^2}$. The detector buried in the glacial ice observes atmospheric muons and neutrinos, secondary particles created by the cosmic rays. The complementary information from the surface and the buried detector allows for a broad range of cosmic ray studies, including mass composition, energy spectra, and muon density in the energy range of 250~TeV to EeV. 

\begin{figure}
    \centering
    \includegraphics[width=0.3\textwidth]{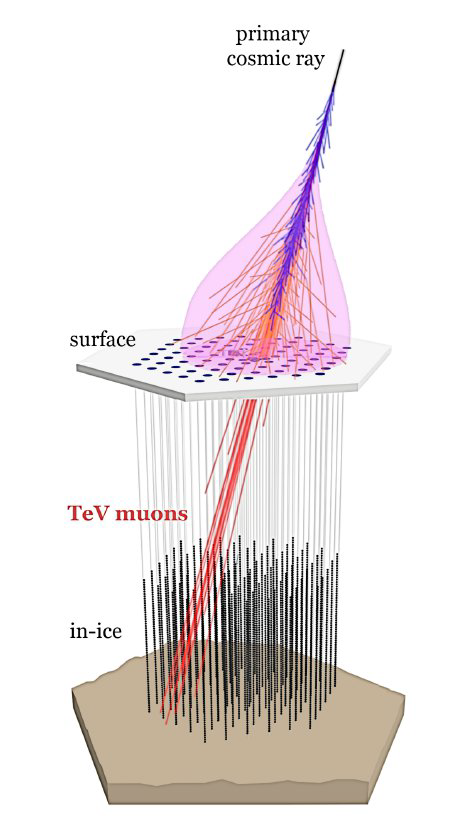}
    \caption{Figure depicts the topology of a coincident detection of near-vertical cosmic ray air shower in IceTop with detectable TeV muons in IceCube.}
    \label{fig:TeVMuons}
\end{figure}

Atmospheric muons trigger the detector at many orders of magnitude higher rate than those of neutrino fluxes. With the enormous data collected, we are able to select a very narrow subset of muons and study the high-energy muon component in near-vertical cosmic-ray air showers detected in coincidence between the surface array IceTop and the in-ice array of IceCube, as shown in Fig.~\ref{fig:TeVMuons}. The combination of the IceTop signal, dominated by the electromagnetic shower component, together with the signal of the muon bundle deep in the ice is used to estimate both the primary cosmic-ray energy and the number of muons with energies above 500 GeV in the shower ("TeV muons"). The average multiplicity of TeV muons is determined for cosmic-ray primary energies between 2.5 PeV and 100 PeV, using three hadronic interaction models: Sibyll 2.1, QGSJet-II.04, and EPOS-LHC. The results are found to be in good agreement with expectations from simulations for all models considered. A possible tension is, however, observed when comparing the results to a low-energy muon measurement performed with IceTop-alone using the models QGSJet-II.04 and EPOS-LHC. These results are of interest in the context of the so-called Muon Puzzle in air-shower physics~\cite{Author:2023icrcTeVmuon}.

Atmospheric neutrinos are detected at a roughly six orders of magnitude lower rate than atmospheric muons. These atmospheric neutrinos are produced in cosmic ray interactions in the atmosphere, mainly by the decay of pions and kaons. With IceCube's extensive data accumulation, we are now able to probe subtle characteristics of atmospheric neutrinos. The rate of neutrinos is affected by seasonal temperature variations in the stratosphere, which are expected to increase with the energy of the particle. Seasonally-dependent energy spectra are obtained for the first time using a novel spectrum unfolding approach, the Dortmund Spectrum Estimation Algorithm (DSEA+), in which the energy distribution from 125GeV to 10 TeV is estimated from measured quantities with machine learning algorithms. The seasonal spectral difference to the annual average flux is determined for the zenith range between $90^{o}$ to $120^{o}$  from 11.5 years of IceCube's upgoing atmospheric muon neutrino data. The zenith region is further restricted to $90^{o}$ to $110^{o}$, as the zenith range from $110^{o}$ to $120^{o}$ shows almost no seasonal variations. The results are compared to the predictions from the numerical cascade equation solver MCEq in this analysis~\cite{Author:2023icrcAtmSeasonal}.

Subdominant to pion and kaon-induced atmospheric neutrinos are the so called "prompt atmospheric neutrinos" produced in decays of charmed mesons that are part of cosmic-ray-induced air showers. Modeling of the production cross section of the heavy mesons is challenging due to poor coverage by collider experiments, and thus the flux of prompt neutrinos is not well known and has not yet been observed by IceCube. This contribution~\cite{Author:2023icrcPrompt} discusses the difficulty of this measurement due to the model dependency of the dominant astrophysical diffuse flux, and how the model assumptions of this dominant flux make upper limits set on the subdominant prompt atmospheric flux unreliable.

Coincidence detection between the surface array IceTop and the in-ice array of IceCube is used for composition analysis of the cosmic ray flux. The ability to detect the electromagnetic and GeV as well as TeV muon content of cosmic-ray-initiated extensive air showers (EAS) probes cosmic ray composition in the energy range of PeV to EeV. Improvements in the composition analysis~\cite{Author:2023icrcComposition} are achieved using air-shower observables sensitive to cosmic-ray primary type and then leveraging a hybrid approach of Graph Neural Network (GNN) to learn hidden correlations in the shower footprint, along with using multiple air-shower observables to capture high-level EAS information.

While IceTop is operating well primarily as a cosmic-ray detector and also as 
a veto for astrophysical neutrino searches for IceCube, the snow accumulation on top of the IceTop detectors increases the detection threshold and attenuates the measured IceTop signals. Enhancing IceTop by a hybrid array of scintillation detectors and radio antennas will lower the energy threshold for air-shower measurements, provide more efficient veto capabilities, enable more accurate cosmic-ray measurements, and improve the detector calibration by compensating for snow accumulation. After the initial commissioning period, a prototype station at the South Pole has been recording air-shower data and has successfully observed coincident events of both the scintillation detectors and the radio antennas with the IceTop array. The production and calibration of the detectors for the full planned array is ongoing. Additionally, one station each has been installed at the Pierre Auger Observatory and the Telescope Array for further R\&D of these detectors in different environmental conditions. Status and plans of these hybrid detector stations for the IceCube Surface Array Enhancement were presented in Ref.~\cite{Author:2023icrcFutureSurface}.

\section{Multi-Messenger Astronomy and Alerts}\label{sec5}
Multi-messenger astronomy is the key driver to the future of high-energy astronomy. Neutrinos have not discovered hidden sources in the electromagnetic spectrum yet, so an a priori source or a model that groups sources needs to be targeted as potential neutrino emitters. The positions of 110 known gamma-ray sources were individually searched for neutrino emission using data between 2011 and 2020. NGC 1068 was observed with a significance of 4.2$\sigma$, which could be associated with neutrino emission from the active galactic nucleus~\cite{ngc1068_paper}. This comes after a 2.9$\sigma$ observation of the same source in~\cite{10yr_paper}. Low level data quality improvements known as ``pass 2'' were carried out as well as improvements in directional reconstructions, and adding 2 years of new data to achieve this observation. 

GRB 221009A is the brightest Gamma Ray Burst (GRB) ever observed. We have used a variety of methods to search for a neutrino counterpart in coincidence with the GRB over several time windows during the precursor, prompt and afterglow phases of the GRB. MeV scale neutrinos are studied using photo-multiplier rate scalers, normally used for supernova searches~\cite{Author:2023icrcSN}. For the first time, a dedicated search for neutrinos below 5~GeV from GRBs was implemented~\cite{Author:2023icrclowEGRB}. These events lack directional localization, but instead can indicate an excess in the rate of events. Neutrinos with energy 10 GeV to TeV and above are searched using traditional methods. The combination of observations by IceCube covers 9 orders of magnitude in neutrino energy, from MeV to PeV, placing upper limits across the range for the allowed neutrino emission~\cite{Author:2023icrcGRB}. The fast-response analysis conducted by IceCube~\cite{GCN_GRB221009} for this GRB in real time shows readiness for IceCube for fast alerts and fast responses in realtime multi-messenger astronomy.

Another alert followup scheme is for gravitational wave events. IceCube follows up LVK alerts sent during O4 using a dataset of high-energy tracks available in low-latency from the South pole~\cite{Author:2023icrcGW}. In fact, the realtime system of IceCube has gone through considerable updates~\cite{icecube_realtime, Author:2023icrcRealtime} and continues to improve with input from our multi-messenger partners. Not only are followup programs being updated as our partners achieve new sensitivities, but we continue to refine our neutrino alert scheme to make them easier for our partners to follow up~\cite{IceCube_GCN, SCRUCD_2023}.

It was following up on our own alert that led to the observation of a neutrino flare of blazar TXS~0506+056~\cite{IceCube:TXS0506}. We searched for additional neutrino emission from the direction of IceCube's highest energy public alert events~\cite{Author:2023icrcAlertFU}. Arrival direction of 122 events with a high probability of being of astrophysical origin were targeted to search for steady and transient emission. In both cases, we find no significant additional neutrino component. The most significant transient emission of all 122 investigated regions remains the flare associated with TXS~0506+056.

Tidal Disruption Events (TDEs) are rare astrophysical transient events that happen when a star passes close to a Supermassive Black Hole (SMBH) that is believed to reside in the center of almost every galaxy. The star can disintegrate, and the resulting stellar debris forms an accretion disk that emits radiation across the electromagnetic spectrum. TDEs have been suggested
as the sources of high-energy neutrinos. We presented a stacking analysis with 29-flare subset of the TDE-like flares using neutrinos with energies above O(100) GeV~\cite{Author:2023icrcTDE}. The result is consistent with background.  

Finally, Active Galacti Nuclei (AGNs) have been a target of IceCube source searches for a long time. While most traditional searches have focused on blazars and AGNs associated with high energy gamma-ray observations, newer searches consider different phase spaces of AGN categories, such as X-ray emission. X-ray bright Seyfert galaxies in the Northern Sky were investigated, both, by assuming a generic single power-law spectrum and spectra predicted by a disk-corona model~\cite{Author:2023icrcSeyfert}. Our results show excesses of neutrinos associated with regions of the sky that contain two sources, NGC 4151 and CGCG 420-015, at a below 3$\sigma$ level. At the same time, this analysis constrains the collective neutrino emission from the source list. Since the AGN environment is rich in gas, dust and photons, they are promising candidate sources of high-energy astrophysical neutrinos, but while the neutrinos manage to escape, the gamma rays may further interact and cascade down to hard X-rays. Therefore, we performed another AGN stacking search and a point source search for high-energy neutrino emission from hard X-ray AGN sampled from the Swift-BAT Spectroscopic Survey (BASS) and presented the results of these two analyses~\cite{Author:2023icrcHardXray}. 

\section{Calibration, Reconstruction, and the IceCube Upgrade}\label{sec6}
In most IceCube analyses, the largest systematic uncertainty is the optical properties of ice. This is because the reconstructions we perform, such as energy, direction, and angular uncertainty, are largely affected by the optical properties of ice. Data rate expectations after cuts can also be very sensitive to these properties. In other words, the better we can calibrate and characterize the local glacial ice properties, the better our analyses become. The importance of this for many IceCube analyses cannot be understated.

To this end, an updated description of the ice was presented~\cite{Author:2023icrcIceModel}. The new description of the ice tilt, which describes the undulation of layers of constant optical properties as a function of depth and transverse position in the detector, has been based on stratigraphy measurements. We now show that it can independently be deduced using calibration data from LEDs buried in the ice at our detector. The new fully volumetric tilt model not only confirms the magnitude of the tilt along the direction orthogonal to the ice flow obtained from prior studies, but also includes a newly discovered tilt component along the flow of ice.

An example of how we implement the constantly improving ice models can be seen in the presentation of the updated HESE event reconstructions~\cite{Author:2023icrcHESEreco}. Several major improvements, including a microscopic description of ice anisotropy arising from ice crystal birefringence and a more complete mapping of the ice layer undulations across the detector, are incorporated into the directional reconstruction of particle showers observed in these high energy cascade events. A reconstruction method that samples posterior distributions across parameters of interest by performing full event resimulation and photon propagation at each step shows an improved per-event description, and updates on previously published source searches using the aggregated sample are presented. 

The IceCube collaboration has even taken the reconstruction challenge public. The Kaggle competition ``IceCube -- Neutrinos in Deep ice'' was a public machine learning challenge designed to encourage the development of innovative solutions to improve the accuracy and efficiency of neutrino event reconstruction. Participants worked with a data set of simulated neutrino events and were tasked with creating a suitable model to predict the direction vector of incoming neutrinos. From January to April 2023, hundreds of teams competed for a total of \$50k prize money, which was awarded to the best performing few out of the many thousand submissions. Insights gained from this project and findings were presented~\cite{Author:2023icrcKaggle}.

\begin{figure}
    \centering
    \includegraphics[width=0.3\textwidth]{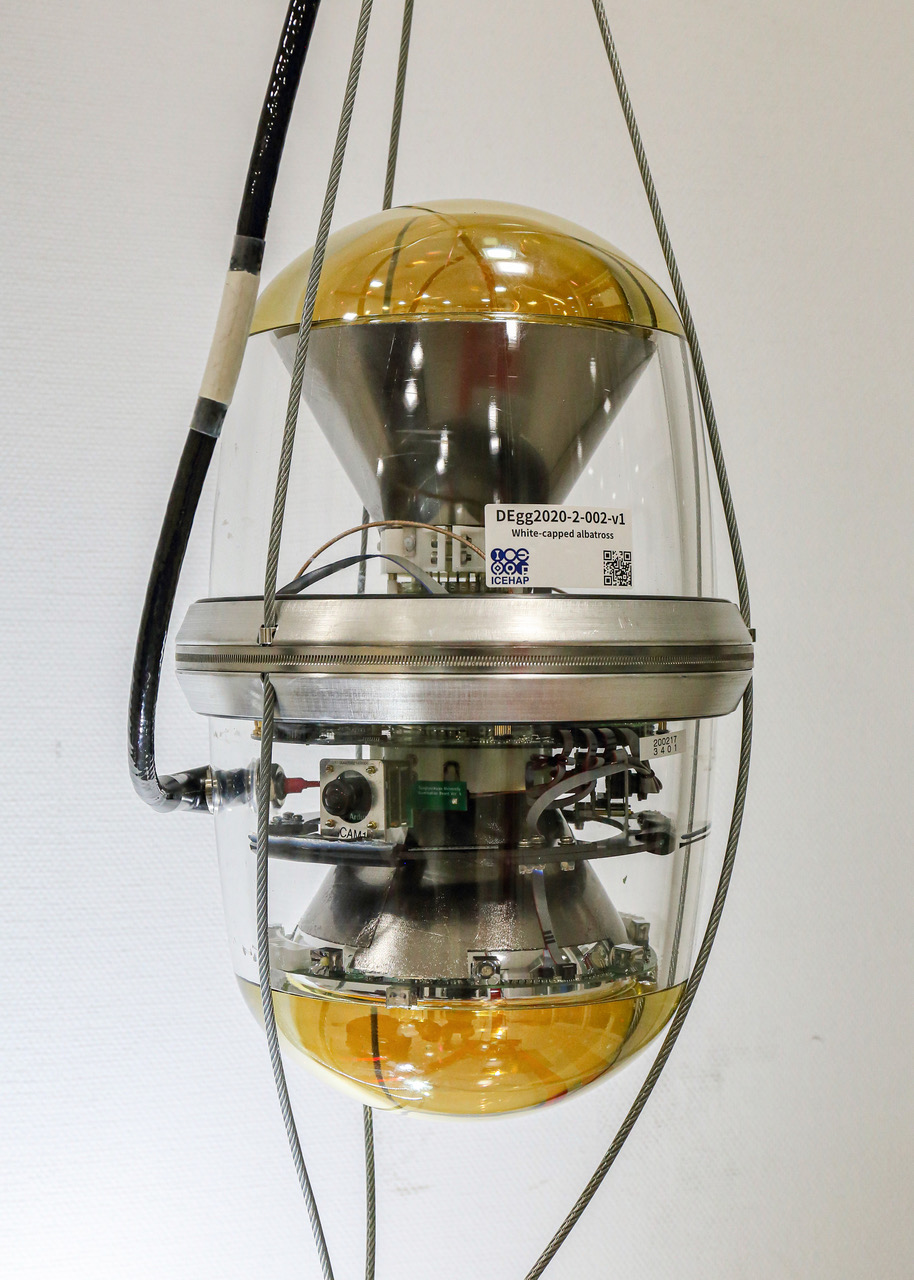}
     \includegraphics[width=0.3\textwidth]{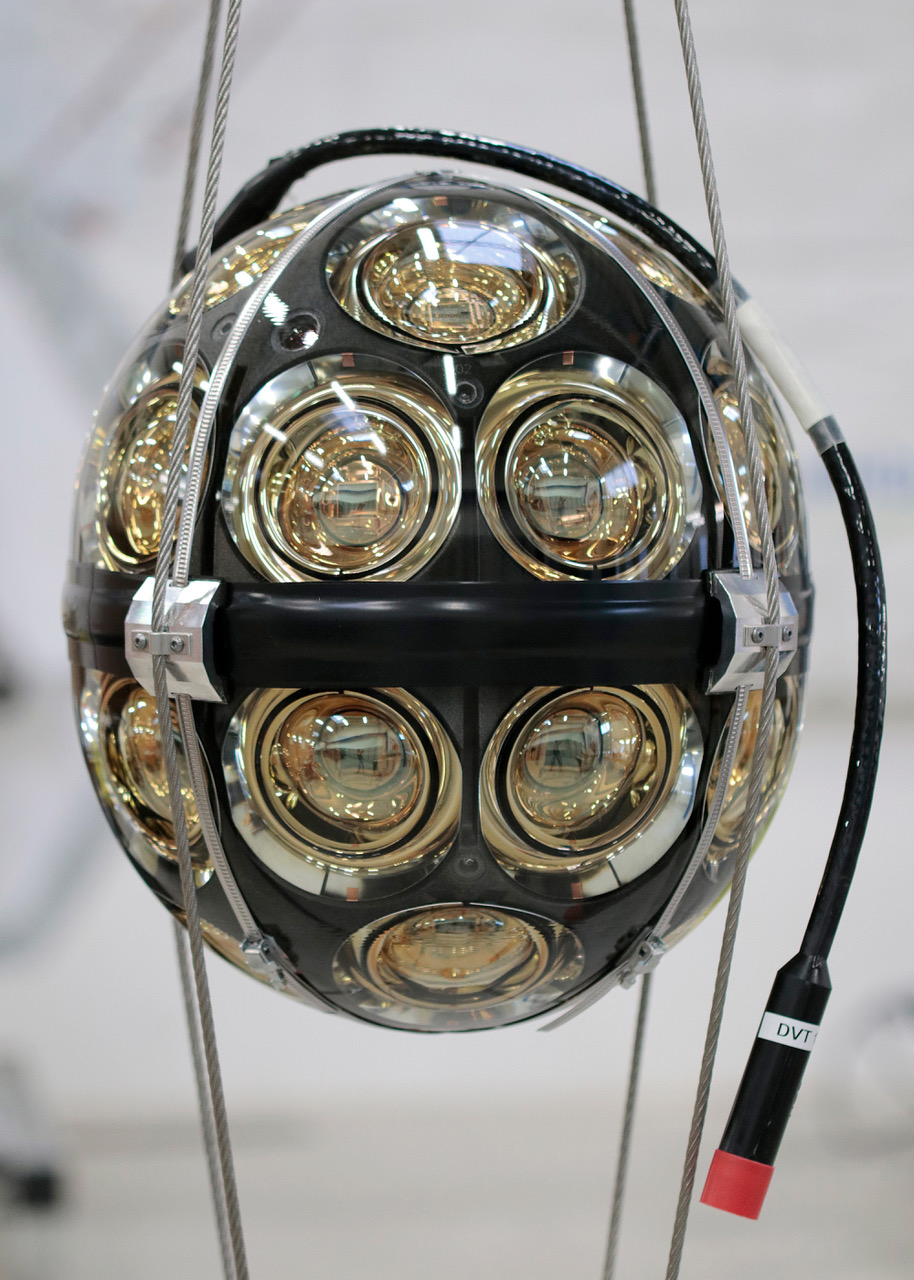}
    \caption{Newly developed multi-PMT optical modules to be deployed in the IceCube Upgrade. Left: D-Egg~\cite{Author:2023icrcDEgg} and Right: mDOM~\cite{Author:2023icrcmDOM}.}
    \label{fig:upgrade}
\end{figure}

A significant opportunity to largely improve our understanding of the optical properties of IceCube's glacial ice is expected in the very near future. A new IceCube extension, called the IceCube Upgrade~\cite{UpgradeICRC}, will be deployed in the polar season of 2025\slash26 and will consist of seven additional strings installed within the DeepCore fiducial volume. The strings will feature new types of optical modules with multi-PMT configurations, shown in Fig.~\ref{fig:upgrade}, as well as new calibration devices. This will significantly enhance our capabilities to calibrate and parameterize the optical properties of the ice. Science goals of the IceCube Upgrade are discussed in the next section.

\section{Other Areas of Study}\label{sec7}
The IceCube detector is a multi-discipline instrument that makes impact in areas beyond those traditionally focused at the Cosmic Ray Conference, such as neutrino oscillations~\cite{Author:2023icrcOsci}, and beyond the standard model physics~\cite{Author:2023icrcQuantumGrav}. Another such area is dark matter physics. Searches can focus on dark matter annihilation and decay at the Galactic Center~\cite{Author:2023icrcDMGC} or one can probe heavier (TeV-PeV) decaying dark matter models. We presented~\cite{Author:2023icrcDM} the first search for neutrinos from dark matter decay in galaxy clusters and galaxies, for dark matter masses ranging from 10 TeV to 1 EeV. Three galaxy clusters, seven dwarf galaxies, and the Andromeda galaxy were selected as targets and stacked within the same source class. No evidence was detected for dark matter decay in the targets. 

With the previously mentioned IceCube Upgrade underway, a significant enhancement of capabilities in the GeV energy range is expected. The new Upgrade strings are expected to increase IceCube’s sensitivity to the atmospheric neutrino oscillation parameters ($\Delta m^{2}_{31}$ and $\Delta m^{2}_{23}$) by about 20-30\%. By using a Graph Neutral Network (GNN), a new analysis framework was developed and studied with simulated IceCube Upgrade data. Uncertainties are approximately half of those using only DeepCore. For the determination of the neutrino mass ordering, IceCube will reach more than $2\sigma$ within a few years of Upgrade operation for any allowed value of $\theta_{23}$ and either mass ordering. Overall, the IceCube Upgrade will significantly improve IceCube’s ability to study atmospheric neutrino oscillations and further improvements are expected when leveraging new calibration that will be deployed with the Upgrade~\cite{Author:2023icrcmUpgradeOsc}, as previously mentioned in \textsection\ref{sec6}.

\section{Conclusion}

IceCube's role as the driver of neutrino astronomy, cosmic ray physics, and areas beyond those was showcased convincingly at ICRC2023. As we enter the era of high statistics data sets, two patterns emerge. One is that using our data wisely in new ways leads to more discoveries. IceCube has always been a data-analysis-driven experiment, and with developments of new data analysis techniques, better reconstruction tools, better understanding of the optical properties of ice, etc. our already-existing decade of data can uncover yet more results to push fields forward. The second trend is that analyses that require extremely rare events, high energy tails of distributions, and very specialized small subsets of data are now becoming statistically feasible to carry out and draw conclusions. These analyses will benefit from even more accumulation of data. In both ways, IceCube is poised to make more discoveries and drive fields forward in the future, including the near future. We look forward to presenting more results at ICRC2025.

\bibliographystyle{ICRC}
\bibliography{references.bib}{}

%

\clearpage

\section*{Full Author List: IceCube Collaboration}

\scriptsize
\noindent
R. Abbasi$^{17}$,
M. Ackermann$^{63}$,
J. Adams$^{18}$,
S. K. Agarwalla$^{40,\: 64}$,
J. A. Aguilar$^{12}$,
M. Ahlers$^{22}$,
J.M. Alameddine$^{23}$,
N. M. Amin$^{44}$,
K. Andeen$^{42}$,
G. Anton$^{26}$
% ...
\\
\\
% ...
$^{12}$ Universit{\'e} Libre de Bruxelles, Science Faculty CP230, B-1050 Brussels, Belgium \\
% ...
$^{17}$ Department of Physics, Loyola University Chicago, Chicago, IL 60660, USA \\
$^{18}$ Dept. of Physics and Astronomy, University of Canterbury, Private Bag 4800, Christchurch, New Zealand \\
% ...
$^{22}$ Niels Bohr Institute, University of Copenhagen, DK-2100 Copenhagen, Denmark \\
$^{23}$ Dept. of Physics, TU Dortmund University, D-44221 Dortmund, Germany \\
% ...
$^{26}$ Erlangen Centre for Astroparticle Physics, Friedrich-Alexander-Universit{\"a}t Erlangen-N{\"u}rnberg, D-91058 Erlangen, Germany \\
% ...
$^{40}$ Dept. of Physics and Wisconsin IceCube Particle Astrophysics Center, University of Wisconsin{\textendash}Madison, Madison, WI 53706, USA \\
% ...
$^{42}$ Department of Physics, Marquette University, Milwaukee, WI, 53201, USA \\
% ...
$^{44}$ Bartol Research Institute and Dept. of Physics and Astronomy, University of Delaware, Newark, DE 19716, USA \\
% ...
$^{63}$ Deutsches Elektronen-Synchrotron DESY, Platanenallee 6, 15738 Zeuthen, Germany  \\
$^{64}$ Institute of Physics, Sachivalaya Marg, Sainik School Post, Bhubaneswar 751005, India \\
% ...

\subsection*{Acknowledgements}

\noindent
The authors gratefully acknowledge the support from the following agencies and institutions:
USA {\textendash} U.S. National Science Foundation-Office of Polar Programs,
U.S. National Science Foundation-Physics Division,
U.S. National Science Foundation-EPSCoR,
Wisconsin Alumni Research Foundation,
Center for High Throughput Computing (CHTC) at the University of Wisconsin{\textendash}Madison,
Open Science Grid (OSG),
Advanced Cyberinfrastructure Coordination Ecosystem: Services {\&} Support (ACCESS),
Frontera computing project at the Texas Advanced Computing Center,
U.S. Department of Energy-National Energy Research Scientific Computing Center,
Particle astrophysics research computing center at the University of Maryland,
Institute for Cyber-Enabled Research at Michigan State University,
and Astroparticle physics computational facility at Marquette University;
Belgium {\textendash} Funds for Scientific Research (FRS-FNRS and FWO),
FWO Odysseus and Big Science programmes,
and Belgian Federal Science Policy Office (Belspo);
Germany {\textendash} Bundesministerium f{\"u}r Bildung und Forschung (BMBF),
Deutsche Forschungsgemeinschaft (DFG),
Helmholtz Alliance for Astroparticle Physics (HAP),
Initiative and Networking Fund of the Helmholtz Association,
Deutsches Elektronen Synchrotron (DESY),
and High Performance Computing cluster of the RWTH Aachen;
Sweden {\textendash} Swedish Research Council,
Swedish Polar Research Secretariat,
Swedish National Infrastructure for Computing (SNIC),
and Knut and Alice Wallenberg Foundation;
European Union {\textendash} EGI Advanced Computing for research;
Australia {\textendash} Australian Research Council;
Canada {\textendash} Natural Sciences and Engineering Research Council of Canada,
Calcul Qu{\'e}bec, Compute Ontario, Canada Foundation for Innovation, WestGrid, and Compute Canada;
Denmark {\textendash} Villum Fonden, Carlsberg Foundation, and European Commission;
New Zealand {\textendash} Marsden Fund;
Japan {\textendash} Japan Society for Promotion of Science (JSPS)
and Institute for Global Prominent Research (IGPR) of Chiba University;
Korea {\textendash} National Research Foundation of Korea (NRF);
Switzerland {\textendash} Swiss National Science Foundation (SNSF);
United Kingdom {\textendash} Department of Physics, University of Oxford.

\end{document}